\documentclass[aps,pre,showpacs,superscriptaddress,twocolumn]{revtex4}
\usepackage{graphicx,epsf}
\usepackage{longtable}%
\begin{document}

\title{Entropy-induced separation of star polymers in porous media}
%
\author{V.~Blavats'ka}
\email[]{viktoria@icmp.lviv.ua}
\affiliation{Institute for Condensed
Matter Physics of the National Academy of Sciences of Ukraine, 
79011 Lviv, Ukraine}
\affiliation{Institut f\"ur Theoretische Physik, Universit\"at Leipzig, 
04109 Leipzig, Germany}
\author{C. von Ferber}
\email[]{ferber@physik.uni-freiburg.de} 
\affiliation{Theoretische Polymerphysik, Universit\"at Freiburg,
79104 Freiburg, Germany}
\affiliation{Complex Systems Research Center, Jagiellonian
University, 31007 Krak\'ow, Poland}
\author{Yu.~Holovatch}
\email[]{hol@icmp.lviv.ua} 
\affiliation{Institute for Condensed
Matter Physics of the National Academy of Sciences of Ukraine,
79011 Lviv, Ukraine} 
\affiliation{Institut f\"ur Theoretische Physik, Johannes Kepler 
Universit\"at Linz, 4040 Linz, Austria}
\affiliation{Ivan Franko National University of Lviv, 
79005 Lviv, Ukraine}
\begin{abstract}
We present a quantitative picture of the separation of star polymers
in a solution where part of the volume is influenced by a porous
medium.  To this end, we study the impact of long-range-correlated
quenched disorder on the entropy and scaling properties of $f$-arm
star polymers in a good solvent. We assume that the disorder is
correlated on the polymer length scale with a power-law decay of the
pair correlation function $g(r)\sim r^{-a}$.  Applying the
field-theoretical renormalization group approach we show in a double
expansion in $\varepsilon=4-d$ and $\delta=4-a$ that there is a range
of correlation strengths $\delta$ for which the disorder changes the
scaling behavior of star polymers. In a second approach we calculate
for fixed space dimension $d=3$ and different values of the
correlation parameter $a$ the corresponding scaling exponents
$\gamma_f$ that govern entropic effects. We find that $\gamma_f-1$, the
deviation of $\gamma_f$ from its mean field value is amplified by the
disorder once we increase $\delta$ beyond a threshold. The consequences
for a solution of diluted chain and star polymers of equal molecular
weight inside a porous medium are: star polymers exert a higher
osmotic pressure than chain polymers and in general higher branched
star polymers are expelled more strongly from the correlated porous
medium. Surprisingly, polymer chains will prefer a stronger correlated
medium to a less or uncorrelated medium of the same density while the
opposite is the case for star polymers.
\end{abstract}
\date{\today}
\maketitle

\section{INTRODUCTION}\label{I}

The influence of structural disorder on the scaling properties of
polymer macromolecules, dissolved in a good solvent is subject to ongoing
 intensive discussions
\cite{Chakrabarti-2005B,Barat-1995I,Harris-1983I,Kim-1983I,Blavatska-2001b,Blavatska-2001a,Blavatska-2002a,Kremer-1981I,Meir-1989I,Nakanishi-1991I,Grassberger-1993I,Ordemann-2000I,Ferber-2003c}.
For polymers,  structural disorder may be realized experimentally by
a porous medium. Depending on the way the latter is prepared, it can
mimic various behavior, ranging from uncorrelated defects
\cite{Li-1992I,Gelb-1998I,Chan-1988I,Yoon-1997I} to complicated fractal objects \cite{Vacher-1988I,Pekala-1995B,Yoon-1998I}.
Consequently, theoretical and Monte Carlo (MC) studies have considered
these different types of
disorder. In particular, the scaling
properties of polymer chains were analyzed for the situations of 
 weak
uncorrelated \cite{Harris-1983I,Kim-1983I}, of long-range-correlated
\cite{Blavatska-2001b,Blavatska-2001a,Blavatska-2002a} as well as of fractal disorder at
the percolation threshold \cite{Kremer-1981I,Meir-1989I,Nakanishi-1991I,Grassberger-1993I,Ordemann-2000I,Ferber-2003c}. However, as far as the authors know,
the influence of correlated disordered media on the
behavior of branched polymers, e.g. polymer stars, have found less attention.
Our work is intended to fill this gap.

The study of star polymers is of great interest since it has a
close relationship to the subject of micellar and other polymeric
surfactant systems \cite{Grest-1996I,Ferber-2002d,Likos-2001I}. Moreover, it can be shown,
 that the scaling behavior of simple star polymers also determines
the behavior of general polymer networks of more complicated structure
\cite{Duplantier-1989I,Schafer-1992}. Recently, progress in
the synthesis of high quality mono-disperse  polymer networks
\cite{Roovers-1972I,Roovers-1983I,Khasat-1988I,Bauer-1989I,Merkle-1993I} has stimulated numerous theoretical studies of
star polymers, both by computer simulation \cite{Grest-1987I,Barrett87,Batoulis-1989I,Ohno-1994I,Shida-1996I,Hsu-2004I} and
by the renormalization group technique
\cite{Duplantier-1989I,Schafer-1992,Miyake-1983I,Duplantier-1986I,Ohno-1988I,Ohno-1989I,Ohno-1991I,Ferber-1995,Ferber-1996b,Ferber-1997c,Ferber-1997d,Ferber-1999,Ferber-2002a,Ferber-2003b,Schulte-2003b}. Let us
note, that polymer stars are hybrids between polymer like entities
and colloidal particles, establishing an important link between
these different systems \cite{Grest-1996I,Ferber-2002d,Likos-2001I,Likos-1998I,Watzlawek-1999I,Jusufi-1999I}.

\begin{figure} [!htb]
\includegraphics[width=6cm]{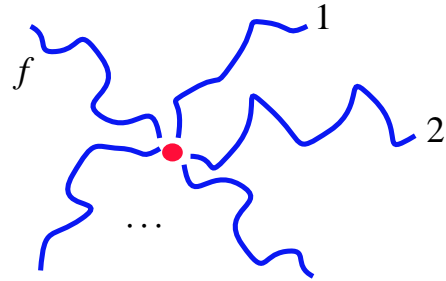}
\caption{\label{fig1}
Polymer star with $f$ arms.}
\end{figure}

It is well established that long flexible polymer chains in good solvents
display universal and self-similar conformational properties on a
coarse-grained scale and that these are perfectly described within a model of
self-avoiding walks (SAWs) on a  regular lattice
\cite{desCloizeaux-1990B,Schafer-1999B,deGennes-1979B}. For the average square end-to-end
distance $R$ and the number of configurations $Z_N$ of a SAW of
$N$ steps one finds in the asymptotic limit $N\to\infty$:
\begin{equation}\label{nu}
\langle R^2 \rangle \sim N^{2\nu^{}},\mbox{\hspace{3em}} Z_N \sim
{\rm e}^{\mu N} N^{\gamma^{}-1}
\end{equation}
where $\nu^{}$ and $\gamma^{}$ are the universal exponents depending
only on the space dimensionality $d$, and ${\rm e}^{\mu}$
 is a non-universal
fugacity.
The universal properties of this polymer model
can be described quantitatively with high precision
by analyzing a corresponding field theory by 
renormalization group methods
\cite{desCloizeaux-1990B,Schafer-1999B,Brezin-1976B,ZinnJustin-1996B,Kleinert-2001B,Amit-1984B}.
 For $d=3$ the exponents read \cite{Guida-1998I} $\nu^{(0)}=0.5882\pm 0.0011$ and
$\gamma^{(0)}=1.1596\pm0.0020$. Here, and in the following we use the notation
$x^{(0)}$ for the value of an exponent derived for the pure solution without
disorder.

The power laws of Eq. (\ref{nu}) can be generalized to describe a star polymer
that consists of $f$ linear polymer chains or SAWs,
linked together at their end-points (see Fig. 1).  For a single star with $f$
arms of $N$ steps (monomers) each,
the number of possible configurations
scales according to \cite{Duplantier-1989I,Schafer-1992}:
\begin{equation}\label{gamstar}
Z_{N,f} \sim {\rm e}^{\mu Nf} N^{\gamma^{}_f-1}\sim(R/\ell )^
{\eta^{}_f-f\eta^{}_2}
\end{equation}
in the asymptotic limit $N\to\infty$ .
The second part shows the power law in terms of the size 
$R\sim N^{\nu^{}}$ of the isolated chain of $N$ monomers on some 
microscopic step length $\ell$, omitting the fugacity factor.  The exponents
$\gamma_f$, $\eta_f$ are universal star exponents, depending on the
number of arms $f$. The relations between these exponents read
\cite{Duplantier-1989I}
\begin{eqnarray} \label{relation}
\gamma_f&=&1+\nu(\eta_{f}-f \eta_2) \nonumber \\
\gamma_1=\gamma_2=\gamma&=&1-\nu\eta_2, \mbox{\hspace{2em}}\eta_1=0.
\end{eqnarray}
Here, $\nu^{}$ and $\gamma^{}$ are usual SAW exponents (\ref{nu}). For
$f=1$, $2$ the case of a single polymer chain is restored. 
Recent numerical values for $\gamma^{}_f$
for different $f$ at $d=3$  are given in Refs.
\cite{Grest-1987I,Barrett87,Batoulis-1989I,Ohno-1994I,Shida-1996I,Hsu-2004I} for Monte Carlo (MC) simulations and
in Refs.\cite{Schafer-1992,Miyake-1983I,Duplantier-1986I,Ohno-1988I,Ohno-1989I,Ohno-1991I,Ferber-1995,Ferber-1996b} for renormalization group 
calculations.

In terms of the mutual interaction, polymer stars interpolate
between single polymer chains (low $f$) and polymeric micelles
(high $f$) \cite{Likos-1998I,Watzlawek-1999I,Jusufi-1999I}.  From the scaling properties of
star polymers, one may also derive their short distance effective interaction.
The mean force $F(r)$ between two star polymers of $f$ and
$f^{\prime}$ arms is inversely proportional to the distance $r<R$ between 
their cores,
\cite{Duplantier-1989I,Ferber-2001a}:
\begin{equation}
\label{force} \frac{1}{k_B T}\,F^{}(r) = \frac{\Theta^{}_{ff'}}{r},
\end{equation}
with the amplitude given by the universal contact exponent 
$\Theta_{ff'}$. The contact
exponents are related to the family of exponents $\eta^{}_f$ for single star
polymers by the following scaling relation  (\ref{gamstar})
\cite{Duplantier-1989I}:
\begin{equation} \label{tetaff}
\Theta^{}_{ff'}=\eta^{}_{f}+\eta^{}_{f'} - \eta^{}_{f+f'}.
\end{equation}

\begin{figure} [!htb]
\begin{picture}(150,200)
\put(-30,30)
{\includegraphics[width=7cm]{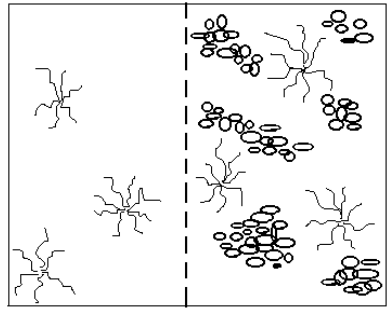}}
\end{picture}
\caption{\label{fig2} Separation phenomena of polymer stars in
good solution, part of which is in a porous medium.}
\end{figure}

Similar to the model of SAWs on a {\em regular} lattice which is used
to describe the scaling properties of long flexible polymer chains in
a good solvent, one may consider models of SAWs on {\em disordered}
lattices to study polymers in a disordered medium. In this model, a
given fraction of the lattice sites is randomly chosen to be forbidden
for the SAW (these forbidden sites will be called defects hereafter).
Harris \cite{Harris-1983I} conjectured that the presence of weak quenched
uncorrelated point-like defects should not alter the SAW critical
exponents. This was later confirmed by renormalization group
considerations \cite{Kim-1983I}.  Another picture appears however for
strong disorder, when the fraction of allowed sites is at the
percolation threshold. Numerous data from exact enumeration,
analytical and MC simulation \cite{Kremer-1981I,Meir-1989I,Nakanishi-1991I,Grassberger-1993I,Ordemann-2000I,Ferber-2003c} strongly suggest that the
scaling of a SAW on a percolation cluster belongs to a new
universality class and is governed by exponents, that differ from
those of a SAW on a regular lattice.

Our present study concerns the scaling properties 
of star polymers in porous media which are found to display
correlations on a mesoscopic scale \cite{Sahimi-1995B}. 
In small angle X-ray and neutron scattering experiments these 
correlations often express themselves by a power law behavior of 
the structure factor $S(q)\sim q^{-d_f}$ on scales 
$\xi^{-1}<q<\ell^{-1}$ where $\ell$ is a microscopic length scale and
$\xi$ is the correlation length of the material and $d_f$ is its
fractal volume dimension \cite{Hasmy-1995I}.
We describe this medium by a model of long-range-correlated 
(extended) quenched defects. This model was proposed
in Ref. \cite{Weinrib-1983I} in the context of magnetic phase
transitions.  It considers defects, characterized by a pair
correlation function $g(r)$, that decays with a distance $r$
according to a power law:
\begin{equation}\label{paircor}
g(r)\sim r^{-a}
\end{equation}
at large $r$. For the structure factor this leads to a power law 
behavior with fractal dimension $d_f=d-a$ where $d$ is the Euclidean
space dimension.
This type of disorder has a direct interpretation
for integer values of $a$. Namely, the case $a=d$ corresponds to
point-like defects, while $a=d-1$ $(a=d-2)$ describes straight lines
(planes) of impurities of random orientation. Non-integer values of $a$ 
are interpreted in terms of impurities organized in fractal 
structures \cite{Weinrib-1983I}.

The influence of the long-range-correlated defects (\ref{paircor})
on magnetic phase transitions has been pointed out in theoretical
work \cite{Weinrib-1983I,Dorogovtsev-1984I,Korutcheva-1998I,Prudnikov-1999Ia,Prudnikov-1999I,Prudnikov-2000I} and MC simulations \cite{Ballesteros-1999I,Vasquez-2003I,Prudnikov-2005I}.
For polymers, its
impact on the scaling of single polymer chains was analyzed in
our previous work in two complementary renormalization group approaches: 
first by a double
expansion in the parameters 
$\varepsilon=4-d$ and the correlation strength $\delta=4-a$
using a linear approximation \cite{Blavatska-2001b} and secondly 
by evaluating two-loop expressions of the theory
for fixed values of $a$ and $d$ \cite{Blavatska-2001a,Blavatska-2002a}. In particular, this
work showed that long-range-correlated disorder leads to a new
universality class with values of the polymer scaling exponents 
that depend on the strength of the correlation expressed by the
 parameters $a$ or $\delta=4-a$. 
From this we may expect that also the architecture dependent scaling
behavior (\ref{gamstar}) of polymer stars and networks is affected
by this type of correlated disorder.

The question we are interested in is: how does the presence of 
long-range-correlated disorder change the values of the critical
exponents (\ref{gamstar}), (\ref{force})? Besides the star-star
interaction, the exponents govern various phenomena that involve star
polymers and polymer networks \cite{Ferber-1997c,Ferber-1997d,Ferber-1999,Ferber-2002a,Ferber-2003b,Schulte-2003b}.
A particular effect that may be observable experimentally 
for star polymer solutions in a porous medium 
is an architecture-dependent impact of the medium on the star polymer.
It may lead to a separation of star polymers with different
numbers $f$ of arms. 
Let us consider star polymers in a good solvent, part of which is in 
a porous medium (see Fig. 2). We consider the pores to be large enough, 
so that the star polymers may pass in and out of the medium
(however, possibly on long time scales only). 
Be ${\cal F}^{(0)}_f(N)$ the free energy of a star polymer with $f$ arms 
of $N$ steps each in the pure solvent 
 and ${\cal F}^{(\delta)}$ its free energy 
in a porous medium characterized by a
correlation strength $\delta$. These can be 
estimated using (\ref{gamstar}):
\begin{eqnarray}\nonumber
{\cal F}^{(0)}_f(N)&=&-\ln {\cal Z}^{(0)}_f(N)\\
&&
=-\mu(0) Nf - 
(\gamma_f^{(0)}-1) \ln N\,, 
\label{osmos1}\\ \nonumber 
{\cal F}^{(\delta)}_f(N)&=&-\ln {\cal Z}^{(\delta)}_f(N)\\
&& =-\mu(\rho) Nf - 
(\gamma_f^{(\delta)}-1) \ln N\,.
\label{osmos2}
\end{eqnarray}
Here, we assume the fugacity factor $e^{-\mu(\rho)}$ to depend on 
the concentration of impurities independent of their correlation,
as it would be the case for SAWs on a lattice with corresponding
defects \cite{Grassberger-1993I}. The product $Nf$ represents the total
number of steps or effective monomers of the star polymer which is
a dimensionless measure of its molecular mass. 
Using Eqs. (\ref{osmos1}) and (\ref{osmos2}) one may now compare
the free energies of a number of situations. Let us name mainly two 
specific questions: (i) Given a star polymer with fixed mass  
$Nf$ and functionality  $f$ in a good solution in a volume
that is influenced by disorder with a fixed defect density $\rho$.
Does the free energy depend on the correlation, and in particular
is the uncorrelated disorder or rather the correlated disorder
of the same density favored by the star polymer? (ii) Given a mixture
of star polymers which are mono-disperse in mass $Nf$ but polydisperse in 
functionality $f$ in a good solution in which only a part of the volume
is influenced by defects (see Fig. \ref{fig2}). 
Due to the fugacity contribution which is the same for all these star
polymers, they are expected to favor the pure part of the solution. However
the extent to which this is the case may depend on architecture.
Is the star polymer mixture partly separated in this situation and
where is the concentration of higher branched star polymers enhanced
in this case? 
While our answer to the first question is mainly to be compared with
MC simulations of star polymers on disordered lattices, the answer to
the second one may also be relevant for experiments with polymers in
solutions inside correlated structures like aerogels.

The setup of the paper is as follows. In the next section we
present the model and construct the  Lagrangean of the 
corresponding field theory. In section III we describe the field-theoretical
renormalization group (RG) methods that we apply. Section IV 
presents our results for the two RG approaches. We conclude with
an interpretation of these results
in section V.

\section {THE MODEL}

Let us consider a single star polymer with $f$ arms immersed in a
good solvent  (Fig. 1). Working within the Edwards
continuous chain model \cite{Edwards-1965I,Edwards-1966I}, we represent
each arm of the star by a path ${\bf r}_{\alpha}(s)$, parameterized by
$0\leq s\leq S$, $\alpha=1,2,\ldots,f$. In a corresponding
discrete model of chains with $N$ steps of mean square 
microscopic length $\ell$ the so-called Gaussian surface is 
$S=N\ell^2$. The central branching point of the star
is fixed at ${\bf r}_1(0)$. 
The partition  function of the system is then defined by the
functional integral \cite{Schafer-1992}:
\begin{eqnarray}\nonumber
{\cal Z}_f(S)&=& \int D [ {\bf r}_1,\ldots, {\bf r}_f ] \\
\label{z} &&\times \exp \left[ -{\cal H}_f
\right]\prod_{\alpha=2}^f\delta^d({\bf r}_\alpha(0)-{\bf r}_1(0)).
\end{eqnarray}
Here, ${\cal  H}_f$ is the Hamiltonian, describing the system of
$f$ disconnected polymer chains:
\begin{eqnarray} \label{eff}
{\cal H}_f&=&\frac{1}{2}\sum_{\alpha=1}^f\int_0^{S}{\rm d}\,s
\left(\frac{{\rm d}\,{\bf r}_{\alpha}(s)}{{\rm d} s}\right)^2\\
&&+\frac{u_0}{4!} \sum_{\alpha,\alpha'=1}^f  \int_0^{S}{\rm d}
s\int_0^{S}{\rm d}s' \delta^d({\bf r}_\alpha(s)-{\bf r}_{\alpha'}(s')).
\nonumber
\end{eqnarray}
The first term in (\ref{eff}) represents the chain connectivity whereas
the second term describes the short range excluded volume interaction. The
product of $\delta$-functions in (\ref{z}) ensures the star-like
configuration of the set of $f$ chains  requiring each of them to start
at the point ${\bf r}_1(0)$. This model may be mapped to a field theory
by a Laplace transformation from the Gaussian surface
$S$ to the conjugated chemical potential variable (mass)
$\hat\mu_0$:
\begin{equation} \label{laplace}
 \widehat{\cal Z}_{f}(\hat\mu_0)=\int{\rm d}S
\exp[-\hat\mu_0 S]{\cal Z}_{f}(S).
\end{equation}
One may then show that the Hamiltonian ${\cal H}$ is related to an
$m$-component field theory with a Lagrangean  ${\cal L}$ 
in the limit $m\to 0$ and that the 
 partition function $ \widehat{\cal Z}_f(\hat\mu_0)$
results from a correlation function of this field theory as follows: 
\begin{widetext}
\begin{eqnarray}
 \widehat{\cal Z}_{f}(\hat\mu_0) &=& \int {\rm d}^d x_1\cdots{\rm d}^d x_f
\langle \sum_{j_1,\ldots,j_f=1}^m
\widehat{T}_{i_1,\ldots,i_f} \phi^{i_1}(x_0)\cdots\phi^{i_f}(x_0)
 \phi^{j_1}(x_1)\cdots\phi^{j_f}(x_f)\rangle^{\cal L}_{m\to 0}\,,
\label{holstar1}\\
{\cal L}&=&\frac{1}{2}
\int{\rm d}^d
x\left[
\right. (\mu_0^2
|\vec{\phi}(x)|^2+|\nabla\vec{\phi}(x)|^2
)+\frac{u_0}{4!}\hat{S}_{i_1,\ldots,i_4}\phi^{i_1}(x)\ldots\phi^{i_4}(x)
\left.\right].\label{holstar}
\end{eqnarray}
\end{widetext}
Here and below, the summation over repeated indices is implied,
$\vec{\phi}$ is an $m$-component vector field
$\vec{\phi}=(\phi^1,\ldots,\phi^m)$, $\hat\mu_0$ and $u_0$ are bare
mass and coupling with the tensor 
$\hat{S}_{i_1,\ldots,i_4}=\frac{1}{3}\left(\delta_{i_1i_2}
\delta_{i_3i_4}+\delta_{i_1i_3}\delta_{i_2i_4}
+\delta_{i_1i_4}\delta_{i_2i_3}\right)$. 
Formally, the local composite operator appearing in Eq. (\ref{holstar1})
is the $m=0$ limit of an operator known in $m$-component field theory 
\cite{Wallace-1975I}:
\begin{equation}
[\phi]^f_*(x) = 
\widehat{T}_{i_1,\ldots,i_f} \phi^{i_1}(x)\ldots\phi^{i_4}(x)
\label{composit}
 \end{equation}
where $\widehat{T}_{i_1,\ldots,i_f}$ is a traceless symmetric $SO(m)$ tensor:
\begin{equation}
\sum_{i=1}^m \widehat{T}_{i,i,i_3,\ldots,i_f}=0.
\label{trace}
\end{equation}

We introduce disorder into the model (\ref{holstar}), by redefining
$\hat\mu_0^2 \to \hat\mu_0^2+\delta\hat\mu_0(x)$,  where the
 local fluctuations
$\delta\hat\mu_0(x)$ obey:
$$
 \langle\langle\delta\hat\mu_0(x)\rangle\rangle=0,
$$ $$
\langle\langle\delta\hat\mu_0(x)\delta\hat\mu_0(y)\rangle\rangle
=g(|x-y|)\,. 
$$ 
Here, $\langle\langle\cdots\rangle\rangle$ denotes
the average over spatially homogeneous and isotropic quenched
disorder. The form of the pair correlation function $g(r)$ is
chosen to decay with distance according to the power law
(\ref{paircor}).

In order to average the free energy over different configurations
of the quenched disorder we apply the replica method to construct
an effective Lagrangean:
\begin{eqnarray}
 {\cal L}_{\rm eff}&=& \frac{1}{2} \sum_{\alpha=1}^{n}
\int{\rm d}^d x [(\hat\mu_0^2|\vec{\phi}_{\alpha}(x)|^2+
|\nabla\vec{\phi}_{\alpha}(x)|^2)
\nonumber\\
&&+\frac{u_0}{4!}
\hat{S}_{i_1,\ldots,i_4}\phi^{i_1}_{\alpha}(x)\ldots\phi^{i_4}_{\alpha}(x)]
\nonumber\\
&&+\sum_{\alpha,\beta=1}^{n}
\int{\rm d}^dx{\rm d}^dy g(|x-y|)
\vec{\phi}_{\alpha}^2(x)\vec{\phi}_{\beta}^2(y).
\label{4}
\end{eqnarray}
Here, the coupling of the replicas is given by  the correlation
function  $g(r)$ of Eq.~(\ref{paircor}), Greek indices denote
replicas and the replica limit $n\to 0$ is implied.

For small $k$, the Fourier-transform $\tilde g(k)$ of $g(r)$
(\ref{paircor}) reads:
\begin{equation}
\tilde g(k)\sim v_0+w_0|k|^{a-d}.
\label{2}
\end{equation}
Thus, rewriting Eq.~(\ref{4}) in momentum space and taking
Eq.~(\ref{2}) into account, one obtains an effective Lagrangean
with three bare couplings $u_0,v_0,w_0$. For $a>d$, the
$w_0$-vertex does not introduce additional divergences at $k=0$
and is irrelevant in the renormalization group sense
\cite{Brezin-1976B,ZinnJustin-1996B,Kleinert-2001B}. The polymer limit $m=0$ leads to further
simplifications. As pointed out in \cite{Kim-1983I}, once the
limit $m,n\to 0$ has been taken, the $u_0$ and $v_0$ terms are of
the same symmetry, and an effective Lagrangean with one coupling
($u_0+v_0$) of $O(mn=0)$ symmetry (\ref{holstar}) results.  This
leads to the conclusion that  weak quenched uncorrelated disorder
i.e. the case $a=d$ is irrelevant for polymers, and consequently 
also for star polymers. 
For $a<d$, the momentum-dependent coupling $w_0k^{a-d}$ has
to be taken into account.  Note that $\tilde g(k)$ must be
positively definite being the Fourier image of the correlation
function. This implies $w_0\geq0$ for small $k$. Also, we assume the coupling
$u_0$ to be positive, otherwise the pure system would undergo a
1st order transition.

The resulting Lagrangean in momentum space then reads:
\begin{widetext}
\begin{eqnarray} \label{h}
&&{\cal L}_{\rm eff}= \frac{1}{2} \sum_{\alpha=1}^n \sum_{k}
(\hat\mu_0^2+k^2)\vec{\phi}^2_{\alpha}(k) +\frac{u_0}{4!}
\sum_{\alpha=1}^n {\sum_{\{k\}}}\delta(k_1{+}\ldots{+}k_4) \vec{\phi}_{\alpha}(k_1)
\cdot \vec{\phi}_{\alpha}(k_2) \vec{\phi}_{\alpha}(k_3) \cdot
\vec{\phi}_{\alpha}(k_4)
\nonumber\\
&+&\frac{w_0}{4!}\sum_{\alpha \beta}^n\! {\sum_{\{k\}}}
\delta(k_1{+}\ldots{+}k_4)
|k_1{+}k_2|^{a-d} \vec{\phi}_{\alpha}(k_1)
\cdot\vec{\phi}_{\alpha}(k_2)\vec{\phi}_{\beta}(k_3) \cdot
\vec{\phi}_{\beta}(k_4).\nonumber
\end{eqnarray}
Here, we have redefined $u_0+v_0\to u_0$ and denoted the
scalar product by
$ \vec{\phi} \cdot \vec{\phi}$.

The replicated composite operator (\ref{composit}) reads in momentum space:
\begin{equation} 
[\phi]^f_*(k_1,\ldots,k_f)=
\delta(k_1{+}\ldots{+}k_f)
\sum_{\alpha=1}^n
\widehat{T}_{i_1,\ldots,i_f}\phi^{i_1}_{\alpha}(k_1)
\ldots\phi^{i_f}_{\alpha}(k_f).
\end{equation}
\end{widetext}

\section {Renormalization group approach}

In order to extract the scaling behavior of
the model (\ref{h}), and of the composite operator (\ref{composit})
we apply the field-theoretical renormalization
group (RG) method \cite{Brezin-1976B,ZinnJustin-1996B,Kleinert-2001B}.  We choose the massive field
theory scheme with renormalization of the one-particle irreducible
vertex functions $\Gamma_0^{(L,N)}(k_1,..,k_L;p_1,..,p_N;\mu^2_0
;\{\lambda_0 \})$ at non-zero mass and zero external momenta
\cite{Parisi-1980I}. The one-particle irreducible (1PI) vertex function
can be defined as:
\begin{widetext}
\begin{eqnarray} \label{vertex} && \delta(\sum k_i + \sum   p_j)
\Gamma_0^{(L,N)}(\{k\};\{p\}; \mu^2_0;\{\lambda_0 \}) =
\int^{\Lambda_0} e^{i(k_i R_i+p_j r_j)} \times\langle
\phi^2(r_1)\dots\phi^2(r_L) \nonumber\\&&\phi(R_1)\dots \phi(R_N)
\rangle^{{\cal  L}_{\rm eff} }_{1PI}
{\rm d}^d R_1 \dots {\rm d}^d R_N {\rm d}^d r_1 \dots {\rm d}^d
r_L \, .
\end{eqnarray}
\end{widetext}
Here,  $\{\lambda_0\}$ stands for the set of  bare couplings 
$u_0,w_0$ of
the effective Lagrangean,  $\{ k\}, \{p \}$ are the sets of
external momenta, $\Lambda_0$ is the cutoff, and the averaging is
performed with the corresponding effective Lagrangean,
 ${\cal L}_{\rm eff}$.
To extract the anomalous dimensions of the composite operators  (\ref{composit})
we define the  additional $f$-point
vertex function $\Gamma^{(f)}_*$, with a single
$[\phi]_*^f$ insertion. Up to 2nd loop order
the graphs for $\Gamma^{(f)}_*$ can be derived from the usual
graphs for $\Gamma^{(0,4)}$ by replacing in turn each four -point vertex
by $[\phi]^f_*$ (see Fig. 3).

\begin{figure} [!htb]
\includegraphics[width=8cm]{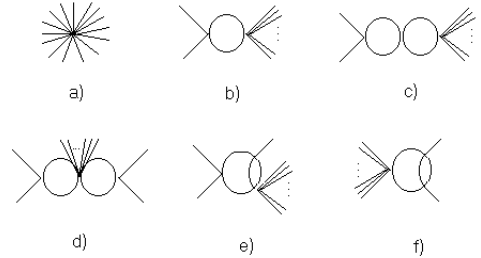}
\caption{The graphs contributing to the vertex function
$\Gamma^{(f)}_*$ up to 2-loop order. (a) represents the $f$-point vertex
$[\phi]^f_*$, (b): one-loop
contribution, (c)-(f): two-loop contributions. \label{diagram}}
 \end{figure}

The
renormalized vertex functions $\Gamma_R^{(L,N)}$ and $\Gamma_{*R}^{(f)}$
are expressed in
terms of the bare vertex functions as follows:
\begin{eqnarray}
&&\Gamma_R^{(L,N)}(\{k\};\{p\};\hat\mu^2;\{\lambda\}) =\nonumber\\&&
Z_{\phi^2}^{L} Z_{\phi}^{N/2}
\Gamma_0^{(L,N)}(\{k\};\{p\};\hat\mu_0^2; \{\lambda_0\}),
\nonumber\\&& \Gamma_{*R}^{(f)}(\{k\};\hat\mu^2;\{\lambda\})
=\nonumber\\&& Z_{*f} \Gamma_{*0}^{(f)}(\{k\};\hat\mu_0^2; \{\lambda_0\}),
\end{eqnarray}
where
$Z_{\phi}$, $Z_{\phi^2}$, $Z_{*f}$ are the renormalizing factors, $\hat\mu$,
$\{\lambda \}$ are the renormalized mass and couplings.

The change of couplings $u_0, w_0$ under
renormalization defines a flow in parametric space, governed by
corresponding $\beta$-functions:
\begin{equation}
\beta_u(u,w)=\frac{\partial u}{\partial \ln
 \ell}\Big|_0,\,\,\beta_w(u,w)=\frac{\partial w}{\partial \ln
 \ell}\Big|_0,\label{betafunc}
\end{equation}
where $l$ is the rescaling factor, and $\Big|_0$   stands for
 evaluation at fixed bare parameters.
 The fixed points (FPs)
$u^*,w^*$ of the RG transformation are
given by the solution of the system of equations:
\begin{equation}
\beta_{u}( u^*,w^*)=0,\phantom{55}\beta_{w}( u^*,w^*)=0\,.
\label{fp}\end{equation}
The stable FP, corresponding to the
critical point of the system, is defined as the fixed point where
the stability matrix
\begin{equation}\label{stmatrix}
B_{ij}=\frac{\partial \beta_{\lambda_i}}{\partial \lambda_j}
\end{equation}
possesses eigenvalues $\{\omega_i\}$ with positive real parts.
The flow of the renormalizing factors $Z_{\phi}$, $Z_{\phi^2}$,
 $Z_{*f}$ in turn defines the corresponding RG functions:
\begin{eqnarray}
 \gamma_{\phi}(u,w)&=&\frac{\partial \ln Z_{\phi}}{\partial \ln
 \ell}\Big|_0, \label{rg1}\\
{\bar \gamma}_{\phi^2}(u,w)&=&-\frac{\partial \ln
 {Z}_{\phi^2}}{\partial \ln \ell }\Big|_0-\gamma_{\phi},
 \label{rg2}\\
 \eta_f(u,w)&=&\frac{\partial \ln Z_{*f}}{\partial \ln
\ell}\Big|_0.\label{rg3}
\end{eqnarray}

The critical exponents are the values of these functions (\ref{rg1})--(\ref{rg3}) at the
stable accessible FP of Eq. (\ref{fp}):
\begin{eqnarray}
&&\nu^{-1}=2-\gamma_{\phi}( u^*,w^*)-
\bar{\gamma}_{\phi^2}( u^*,w^*),\\
&& \eta =\gamma_{\phi}( u^*,w^*),\label{eta}\\
&& \eta_f =\eta_{f}( u^*,w^*),\label{eta_f}\\
&& \gamma_{f}=1+\nu\eta_f(u^*,w^*)+(\nu(2-\eta)-1)f.\label{gammaf}
\end{eqnarray}
Here, $\eta_f$ is the anomalous dimension of the composite
operator $[\phi]^f_*$. The expressions for the
exponents $\nu$, $\eta$ of a single polymer chain in
long-range-correlated disorder we derived in 
Ref. \cite{Blavatska-2001b,Blavatska-2001a,Blavatska-2002a}. Only the RG functions
$\eta_f$ that correspond to the anomalous dimensions of the composite
operator $[\phi]^f_*$ in the presence of correlated disorder 
remain to be calculated in order to extract the
spectrum of star polymer exponents $\gamma_f$ (given by Eq.
(\ref{gammaf})).

\section{The results}

The perturbative expansions for the functions (\ref{betafunc}),
(\ref{rg1}) - (\ref{rg3}) may be analyzed by two complementary
approaches: either by exploiting a double expansion in
$\varepsilon=4-d,\delta=4-a$ \cite{Weinrib-1983I,Dorogovtsev-1984I,Korutcheva-1998I,Blavatska-2001b}
or by evaluating the theory for fixed values of the parameters $d$ and $a$
\cite{Prudnikov-1999Ia,Prudnikov-1999I,Prudnikov-2000I,Blavatska-2001a,Blavatska-2002a}. In the following we make use of both ways
of analysis.

\subsection{One-loop approximation:
$\varepsilon,\delta$ - expansion}

For the qualitative analysis of the first order results, we apply
a double expansion in $\varepsilon=4-d$ and $\delta=4-a$. First, we need
to calculate the $f$-point vertex function $\Gamma^{(f)}_*$ with
a single
insertion of the composite operator $[\phi]_*^f$. In the
one-loop approximation we get:
\begin{eqnarray}
&&\Gamma^{(f)}_*(u_0,w_0,\{k\}=0,\hat\mu_0)=1-u_0\,\frac{f(f-1)}{6}
\nonumber\\&& \times \int\frac{{\rm d}{\vec q}}{(q^2+\hat\mu_0^2)^2}
+w_0\,\frac{f(f-1)}{6}\int\frac{{\rm d}{\vec
q}\,q^{a-d}}{(q^2+\hat\mu_0^2)^2}. \label{baregf}
\end{eqnarray}
We define renormalized mass $\hat\mu^2$ and couplings $u,v$ by the
renormalization conditions:
\begin{eqnarray}
\hat\mu^2&=& \Gamma_{R}^{(2)}(k,\hat\mu^2,u,w)|_{k=0}, \nonumber\\
 u &=&\Gamma_{R,u}^{(4)}(\{k\},\hat\mu^2,u,w)|_{\{k\}=0},  \nonumber\\
 w &=&\Gamma_{R,w}^{(4)}(\{k\},\hat\mu^2,u,w)|_{\{k\}=0}. \nonumber
\end{eqnarray}
The renormalization condition for the vertex function with
$[\phi]_*^f$-insertion is given by
 \begin{eqnarray} \nonumber
Z_{*f}(u,w)&=&[\Gamma^{(f)}_*(u,w)]^{-1} \\ &&
=1+
u\frac{f(f-1)}{6}I_1-w\frac{f(f-1)}{6}I_2\,.
\label{zf}
 \end{eqnarray}
Here, $I_1$ and $I_2$ are loop integrals given in the
appendix. The expressions for the RG $\beta$- and
$\gamma$-functions (\ref{betafunc}), (\ref{rg1}), (\ref{rg2})
within the same approximation read \cite{Blavatska-2001b}:
\begin{eqnarray}  \nonumber
\beta_u &=& - \varepsilon \left[ u- \frac{4}{3}\,u^2 I_1 \right ]
-\delta 2uw \left[ I_2-\frac{1}{3} D_1 \right]
\\&+&  \label{beta1laa}
 (2 \delta-
\varepsilon) \frac{2}{3}\,w^2 I_3,\\ \label{beta1la} \beta_w &=&
-\delta \left [w + \frac{2}{3}\,w^2 I_2+ w^2 D_1 \right ]
+\varepsilon \frac{2}{3} \left[wu I_1 \right],\\
&&\gamma_{\phi^2}= \varepsilon
\frac{u}{3}\, I_1 -\delta \,\frac{w}{3}\,I_2 , \phantom{55}
\gamma_{\phi}=\delta\, \frac{w}{3} D_1.  \label{B}
\end{eqnarray}
Again, the loop integrals $I_1-I_3, D_1$ are given in the
Appendix. Note that contrary to the usual $\phi^4$ theory the
$\gamma_{\phi}$ function in Eq.~(\ref{B}) is nonzero already in
the one-loop order. This is due to the $k$-dependence of the
integral $D_1$. Combining Eqs. (\ref{rg3}) and (\ref{betafunc})
one defines $\eta_f$ via familiar expressions (\ref{zf}) and
(\ref{beta1la}):
\begin{equation}
\eta_f=\beta_u(u,w)\frac{\partial \ln Z_{*f}}{\partial
u}+\beta_w(u,w)\frac{\partial \ln Z_{*f}}{\partial w}.\label{eta_ff}
\end{equation}

\begin{figure} [!htb]
\includegraphics[width=8cm]{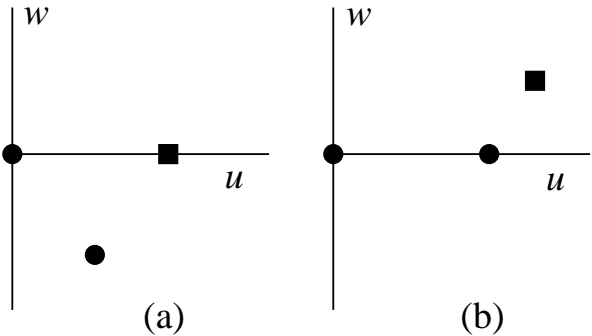}
\caption{Fixed point picture for $d<4$, $a>d$.
Stable physically accessible fixed points are shown by squares
the unstable ones by discs.
 (a): $\delta<\varepsilon/2$, the pure fixed point ($u^*\neq 0,w^*=0$) is stable. At
$\delta = \varepsilon/2$, it interchanges its stability with the
LR fixed point ($u^*\neq 0,w^*\neq 0$). (b): for
$\varepsilon/2<\delta<\varepsilon$ the LR fixed point becomes physically
accessible and stable. The Gaussian fixed point ($u^*=0,w^*=0$) is stable for
$d>4,a>4$. \label{points} }
 \end{figure}

To proceed with the analysis, we insert the expansions of
the one-loop integrals:
\begin{eqnarray} \label{eps1}
I_1&=& \frac{1}{\varepsilon}\left (1 -
 \frac{\varepsilon}{2} \right),\\ \label{eps2}
 I_2 &=&
 \frac{1}{\delta}\left
 (1 - \frac{\delta}{2} \right),\\ \label{eps3}
I_3 &=&  \frac{1}{2\delta-\varepsilon}\left
 (1 - \frac{2\delta-\varepsilon}{2} \right),\\ \label{eps4}
D_1 &=& \frac{1}{\delta}\left
 (\frac{\delta-\varepsilon}{2} \right).
\end{eqnarray}
Substituting (\ref{eps1}) - (\ref{eps4}) into the expressions for
$\beta$-functions (\ref{beta1laa}), (\ref{beta1la}) and solving
the FP equation (\ref{fp}), one finds three fixed points: the
Gaussian ($u^*=0,w^*=0$), the pure
($u^*=\frac{3}{4}\varepsilon,w^*=0)$, and the non-trivial,
long-range-correlated, LR fixed point:
($u^*=\frac{3}{4}\frac{2\delta^2}{\varepsilon-\delta},
w^*=\frac{3}{2}\frac{\delta(\varepsilon-2\delta)}{\delta-\varepsilon}$).
The analysis of the conditions of their stability and accessibility
we performed in Ref. \cite{Blavatska-2001b}. The results are displayed
schematically in Fig.
{\ref{points}}: at $\varepsilon, \delta>0$, the crossover from the
pure FP to the LR takes place at $\delta=\varepsilon/2,$ i.e.
$a=2+d/2$. Note, however, that the LR FP is stable in the region $a>d$, where
the influence of the disorder is expected to be irrelevant,
see the explanation after
Eq. (\ref{2}).  These first order results give a qualitative
description of the crossover to the new universality class in the
presence of long-range-correlated disorder.

The expression for the critical exponent $\nu$ reads
\cite{Blavatska-2001b}:
\begin{eqnarray}
\nu &=& \left \{ \begin{array}{ll} \nu^{(0)}=
1/2 + \varepsilon/16,\; & \delta<\varepsilon/2, \\
\nu^{(\delta)}= 1/2 + \delta/8,\; & \varepsilon/2
<\delta<\varepsilon.
\end{array}\right.\label{nu1}
\end{eqnarray}
From (\ref{eta_ff}) we find:
\begin{eqnarray}
\eta_f &=& \left \{ \begin{array}{ll} \eta_f^{(0)}=
-\frac{1}{8}\varepsilon f(f-1),\; & \delta<\varepsilon/2, \\
\eta_f^{(\delta)}= -\frac{1}{4}\delta f(f-1),\; & \varepsilon/2
<\delta<\varepsilon,
\end{array}\right.\label{eff1}
\end{eqnarray}

Substituting (\ref{eff1}), (\ref{nu1}) into (\ref{gammaf}),
finally we get:
\begin{eqnarray}
\gamma_f= \left \{ \begin{array}{ll} \gamma_f^{(0)}=
1-\frac{1}{16}\varepsilon f(f-3) \; & \delta<\varepsilon/2, \\
\gamma_f^{(\delta)}= 1-\frac{1}{8}\delta f (f-3)\; & \varepsilon/2
<\delta<\varepsilon,
\end{array}\right.\label{glr}
\end{eqnarray}
The first line in (\ref{glr}) recovers the exponent for the
$f$-arm polymer star in the pure solution \cite{Miyake-1983I}, whereas the second line
brings about a new scaling law.

To obtain a naive estimate of the numerical values of these
exponents, one can directly substitute into
 (\ref{glr}) the value $\varepsilon=1$ (corresponding to $d=3$) and
 different fixed values for correlation parameter $a$ 
 We note a decrease of the star exponent $\gamma_f$ at
 fixed $f>3$, when the correlation of the disorder becomes stronger
 (i.e. parameter $a$ decreases). However, the behavior for chain 
polymers i.e. for $f=1,2$ differs: in this case the exponents 
$\gamma_1=\gamma_2$ increase for decreasing $a$.

 \begin{figure} [!htb]
\includegraphics[width=8cm]{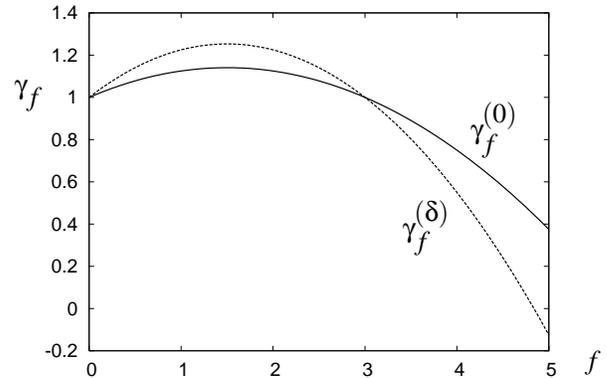}
 \caption{\label{fig-loop1} 
 Comparison of the exponents $\gamma_f$ for star polymers in 
 a pure solution (continuous line) and in a solution inside a correlated
 porous medium (broken line), following Eq.(\ref{glr}) for 
 $\varepsilon=1$, $\delta=0.9$. }
 \end{figure}

This crossover is also clearly seen in Fig. \ref{fig-loop1} where
we compare the behavior of $\gamma_f$ for the case with and without
correlated disorder.As this figure shows, the correlation of the 
disorder effectively enhances the deviation from the mean field
value $\gamma_f^{\rm MF}=1$ which is positive for $f=1,2$ and negative
for $f>3$ in this approximation.

\subsection{Two-loop approximation: fixed $d,a$ approach}

To obtain a quantitative description of the scaling behavior of star polymers
in long-range-correlated disorder, we proceed to higher order
approximations. We make use of the fixed $d=3$ RG approach
\cite{Parisi-1980I}, considering the massive RG functions at fixed
space dimension $d$. Also the additional parameter $a$ in the
expansions for the RG functions  in renormalized couplings $u$, $w$ 
(\ref{betafunc}),
(\ref{rg1})-(\ref{rg3}) is fixed in this approach and we work hereafter with
these expansions. 
As is well known
\cite{Brezin-1976B,ZinnJustin-1996B,Kleinert-2001B}, such expansions 
are in general characterized by a
factorial growth of the coefficients which implies a zero radius of
convergence \cite{Hardy-1948B}.  No reliable data 
can be extracted from a naive analysis. 
For the present model, this particular
feature shows up already in the first order of perturbation theory in
$u$ and $w$. Indeed, for the plain one-loop $\beta$-functions (\ref{beta1la})
a non-trivial FP LR does not appear if one solves the 
non-linear fixed
point equation (\ref{fp}) directly at $d=3$ and $2<a<3$. To take into
account higher order contributions, the standard
tools of asymptotic series resummation have to be applied \cite{Hardy-1948B}.

The two-variable Pad\'e-Borel resummation technique \cite{Jug-1983I}
that we use consists of several steps. Consider the two-variable series
for a RG function $h(u,w)$. First, we construct the Borel
image of the initial function: 
$$
h(u,w)=\sum_{i,j}a_{i,j}u^iw^j\to\sum_{i,j}\frac{a_{i,j}(ut)^i(w
t)^j}{\Gamma(i+j+1)}, $$ where $\Gamma(i+j+1)$ is Euler's gamma
function. Then, the Borel image is extrapolated by a rational
Pad\'e approximant $[K/L](u,w)$. This ratio of two
polynomials of order  $K$ and $L$ is constructed as to match its truncated Taylor
expansion to that of the Borel image of the function $h(u,w)$.
The resummed function is then recovered by an inverse Borel
transform of this approximant:
\begin{equation}
h^{\rm res}(u,w)=\int_0^{\infty}{\rm d}t \exp(-t)[K/L](ut,wt).
\end{equation}

In our previous work \cite{Blavatska-2001a,Blavatska-2002a} we have analyzed the
resummed expressions for the two-loop RG functions of the model of
a single polymer chain in the long-range-correlated disorder in
three dimensions, and found that a fixed point LR  appears
and is stable at $2.2<a<3$. This FP disappears at $a<2.2$ and the
pure SAW FP remains unstable. This behavior may be interpreted 
to indicate, that the presence of stronger correlated
disorder (at $a<2.2$) might lead to a collapse of the polymer chain. To
obtain a quantitative picture of the scaling behavior  of star
polymers, we only need to extend these results by 
a calculation of the
renormalization factor $Z_{*f}$ (\ref{eta_ff}). Taking into account
the two-loop contributions shown in Fig. \ref{diagram} we get:
\begin{widetext}
\begin{eqnarray}
Z_{*f}&=&1+u\,\frac{f(f-1)}{6} I_1- w\,\frac{f(f-1)} {6} I_2\nonumber\\&&
+u^2\,f(f-1) \left[ \left(-\frac{1}{72}(f-2)(f-3) \right.\right.
\left.+\frac {1}{36}f(f-1)+\frac{1}{6}\right)I_1^2\nonumber\\
&&-\left(\frac{f-2}{9} +\frac{1}{6} \right) I_6\Big]
-uw\,f(f-1)\left[\left(-\frac{1}{36}(f-2)(f-3)+\frac{1}{18}f(f-1)
\right.\right.\left.+ \frac{7}{8}\right)I_1I_2\nonumber\\
&&+\left(-\frac{1}{9}(f-2)-\frac{1}{3}\right)I_7
-\left.\frac{1}{9}(f-2)I_9-\frac{I_4}{18}-\frac{2}{3
f(f-1)}D_1\right]\nonumber\\&&
 +w^2\,f(f-1)\left[\left(-\frac{(f-2)(f-3)}{72}
-\frac{f(f-1)}{36}\right.\right. \nonumber\\ \label{res1}
&&+\frac{1}{9}\left.\right)I_2^2-\left(\frac{f-2}{18}
+\frac{1}{6}\right)I_8- \frac{f-2}{18}I_{10}+\frac{1}{9}I_3I_1
-\left.\frac{1}{18}I_5-\frac{2}{3f(f-1)}I_2D_1\right].
\end{eqnarray}
\end{widetext}
The expressions for the loop integrals $I_1,\ldots, I_{10}, D_1$
and their numerical values at $d=3$ and different $a$ are
presented in the Appendix.

In this way, the function $\eta_f$ can be found, using (\ref{eta_ff}) and
familiar expressions for the two-loop $\beta$-functions as given in
\cite{Blavatska-2001a,Blavatska-2002a}. The resulting two-loop expansion  for
$\eta_f$ reads \cite{note2}:
\begin{widetext}
\begin{eqnarray}
\eta_f&=&-u\frac{f(f-1)}{8}-w\frac{(4-a)
f(f-1)}{8}I_2/I_1
\nonumber\\&&
+u^2 \left(\left(-\frac{f^3}{16}+\frac{3f^2}{32}-
\frac{f}{32}\right)I_1^2 + \left( \frac
{f^3}{8}-\frac{3}{16}f^2+\frac{f}{16} \right)I_6
      \right)\Big/I_1^2 \label{res}\\
&&+uw \Big( \left( -\frac{f^3}{16}(1+(4-a))+
 \frac{f}{16}(1-(4-a))+(4-a)\frac{f^2}{8}\right)I_1I_2
\nonumber\\&&
+
 \left((1+(4-a))\left(\frac{f^3}
{16}-\frac{3f^2}{32}+\frac{f}{32}\right)\right)(I_7+I_9)
\nonumber\\ &&
+\left((1+(4-a))\left(\frac{f^2}{32}-\frac{f}{32}\right)\right)I_4\nonumber\\&&
+
\left(\frac{3}{8}(1+(4-a))\right.
-\left.\frac{f^2(4-a)}{16}
+\frac{(4-a)f}{16} \right)\frac{I_1D_1}{f(f-1)}
)\Big/I_1^2 \nonumber\\&&
+w^2\Big(
\left(-\frac{f^2}{16}-\frac{f}{16}\right)I_1I_3
+\frac{(4-a)f(f-1)}{16}I_5+
\nonumber\\&&
+
\frac{(4-a)(f-1)(f+1)}{16}I_8+\frac{(4-a)f(f-1)(f-2)}{16}I_{10}
\nonumber\\&&+\left((4-a)-\frac{f^2}{16}+
\frac{f}{6}+\frac{3}{4}\right)I_2D_1\Big)\Big/I_1^2\nonumber\,.
\end{eqnarray}
\end{widetext}

Inserting the series for $\nu$ and $\eta$ for the polymer chain
in the long-range-correlated disorder from 
Refs. \cite{Blavatska-2001a,Blavatska-2002a}
together with $\eta_f$ of Eq. (\ref{res}) into
(\ref{gammaf}), we finally obtain the corresponding series for $\gamma_f$. Substituting
the numerical values of the LR correlated FP, found for different
$a$ from Refs. \cite{Blavatska-2001a,Blavatska-2002a} and applying
a Pad\'e-Borel resummation as explained above we get the numerical values for exponents
$\gamma_f$ in three dimensions for different values of the
correlation parameter $a$ and number of arms $f$. Our final 
estimates that result from this procedure are presented in Table 
\ref{result}. For $f=3$, the
first order contribution to $\gamma_f$ is zero, whereas it is non-zero
for $\eta_3$. Therefore, to obtain a resummed value for $\gamma_3$ 
we have resummed the series for $\eta_3$ using the values for $\eta$
and $\nu$ of chain polymers in long-range-correlated disorder.
\begin{widetext}
\begin{center}
\begin{table}[ht!]
\begin{tabular}{|l| llllllll   |} \hline $a \setminus f$ & 1;2 & 3 &
4 &5 & 6 &7 & 8 & 9 \\ \hline
 3, \cite{Ferber-1995,Ferber-1996b} &
1.18 & 1.06 & 0.86 & 0.61 & 0.32 & -0.02 & -0.4 & -0.8
\\
3, \cite{Hsu-2004I}  & 1.1573(2) &1.0426(7)&
0.8355(10)
&0.5440(12)&0.1801(20)&-0.2520(25)&-0.748(3)&-1.306(5)\\
3& 1.17& 0.99&0.83&0.57&0.26&-0.08&-0.56&-0.87\\
 2.9 & 1.25 &0.87 & 0.78 & 0.46  & 0.09  &
-0.32 & -0.76 &
-1.23 \\
2.8 & 1.26 & 0.81& 0.76 & 0.43  & 0.06  & -0.36  & -0.80
 & -1.26 \\
 2.7&   1.28 & 0.74& 0.72 & 0.40
  & 0.01  & -0.40  & -0.85 & -1.31 \\
 2.6 & 1.30 & 0.73&  0.70
  & 0.37  & -0.03 & -0.46 & -0.91 & -1.37 \\
2.5 & 1.34&  0.71 & 0.70 & 0.35
  & -0.10 & -0.51 & -1.00 & -1.44 \\
 2.4 & 1.35 & 0.70&  0.70
  & 0.31  & -0.10 & -0.55 & -1.02 & -1.50 \\
2.3 & 1.38 & 0.70&  0.69
 & 0.29  & -0.13 & -0.59 & -1.06 & -1.55 \\
\hline
\end{tabular}
\caption{Critical exponents $\gamma_f$ for the $f$-armed star in
three dimensions at different values of the correlation parameter
$a$. The first and the second rows ($a=d=3$) present results for a
polymer star in a good solvent without porous medium obtained
within the field-theoretical RG in three-loop approximation, Ref.
\cite{Ferber-1995,Ferber-1996b}, and by the Monte Carlo simulations, Ref.
\cite{Hsu-2004I}, correspondingly. \label{result}}
\end{table}
\end{center}
\end{widetext}

Let us recall, that for $a=d=3$ the problem 
is equivalent to the situation without
 structural
disorder. Therefore, in the first two rows of Table \ref{result}
we give RG estimates for the exponents $\gamma_f$ obtained
in a three-loop approximation in Ref. \cite{Ferber-1995,Ferber-1996b} as well as
recent data of MC simulations \cite{Hsu-2004I}. Comparing 
these data with our two-loop results (the third row of the Table)
allows to estimate the consistency of the calculational scheme
that we apply. The good mutual agreement found for
the low values of $f$ supports our approach. 
The fact that the discrepancy increases with $f$ is expected, 
taking into account the strong combinatorial
$f$-dependence of the coefficients of expansions (\ref{res1}),
(\ref{res}). This growth is difficult to control in a consistent
way during the resummation.

As we noted above, the choice $f=1,f=2$ recovers the case of a
single polymer chain. Therefore, the first and the second columns
of Table 1 give an estimate for the dependence of the exponent
$\gamma$, Eq. (1): $\gamma^{(a)}=\gamma_1^{(a)}=\gamma_2^{(a)}$. The
remarkable feature of the estimates for $\gamma_f^{(a)}$ listed in
the Table 1 is that they predict a qualitatively different behavior
of $\gamma_f^{(a)}$ for $f=1,f=2$ and $ f\geq 3$. Indeed, as one
sees from Table 1, a decrease in $a$ leads to an increase 
of $\gamma_1^{(a)},\gamma_2^{(a)}$ while $\gamma_f^{(a)}$ for 
$f\geq 3$ decrease in this case. This
tendency is also found for the one-loop $\varepsilon,\delta$-expansion (Eq.
(\ref{glr})).

Recall, that the scaling exponent of a star polymer in a pure
solvent is given by $\gamma_f^{(a=3)}$ and let us return back 
to Eqs.
(\ref{osmos1}) and (\ref{osmos2}) for the free energy of a star in
the pure solvent and in a porous medium.
Then our results indicate two
different regimes of the entropy-induced change of the polymer
concentration for a solvent in a porous medium with respect to the
pure one. Namely, the free energy of the chain polymers
($f=1,f=2$) is reduced by the presence of correlation in a 
porous medium. 
On the other hand, the free energy of a star polymer
($f\geq 3$) is increased by correlations of the environment.

\begin{table}[ht!]
\begin{center}
\begin
{tabular}{|c|  c c c  c  c  c c  c|}\hline
$a$ & $\Theta_{41}$  & $\Theta_{43} $& $\Theta_{44}$  &  $
\Theta_{45} $ & $ \Theta_{46} $
  & $ \Theta_{47} $
& $ \Theta_{48} $ & $ \Theta_{49} $  \\ \hline
2.9 & 1.306 &1.627 & 1.865 & 2.042 & 2.220 & 2.348 & 2.402 & 2.531 \\
2.8  & 1.286& 1.565&  1.777 & 1.932 & 2.071 & 2.163 & 2.246 & 2.345\\
2.7 & 1.262 &  1.502& 1.691 & 1.817 & 1.941 & 2.024 & 2.082 & 2.166\\
2.6& 1.239&1.459 & 1.608 & 1.739 & 1.843 & 1.910 & 1.987 & 2.040\\
2.5& 1.229& 1.410& 1.554 & 1.668 & 1.762 & 1.834 & 1.876 & 1.929\\
2.4& 1.217&  1.392& 1.521 & 1.617 & 1.705 & 1.758 & 1.799 & 1.883\\
2.3& 1.193&1.360 & 1.474 & 1.574 & 1.651 & 1.707 & 1.725 & 1.772\\
\hline
\end{tabular}\end{center}
\caption{Contact exponents $\Theta_{ff}$, governing the scaling behavior of
interaction force between two $f$- and $f'$-armed polymer star in three
dimensions at different values of the correlation parameter $a$.
\label{teta}}
\end{table}

\begin{figure} [!htb]
\includegraphics[width=8cm]{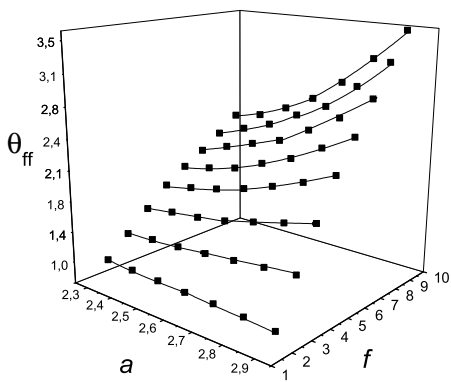}
\caption{The contact exponent $\Theta_{ff}$ as function of $f$ and
correlation parameter $a$ at $d=3$. Each line shows the
dependence of $\Theta_{ff}$ on $a$ at fixed $f$. }
 \label{tetaf}
 \end{figure}

\begin{figure} [!htb]
\includegraphics[width=8cm]{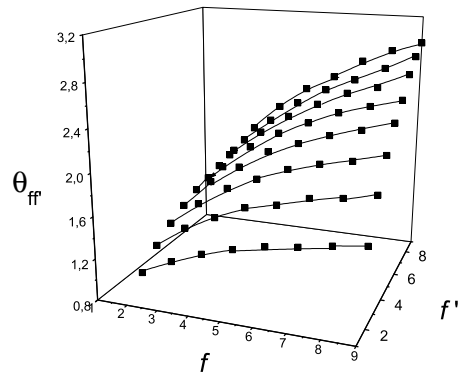}
\caption{The contact exponent $\Theta_{ff'}$ as function of $f$
and $f'$ for fixed correlation parameter $a=2.7$ at $d=3$.
\label{fig-tetaff}}
 \end{figure}

To investigate the influence of a porous medium on the effective interactions between star polymers 
we calculate the contact exponents $\Theta_{ff'}$ (Eq. \ref{force}).
Our results, obtained by a
Pad\'e-Borel resummation of the series derived from Eq. (\ref{tetaff}). 
are presented in
Figs. \ref{tetaf}, \ref{fig-tetaff} and for a selected set of exponents
also  in Table \ref{teta}.
In Fig. 5 we show the contact
exponent $\Theta_{ff}$ for two stars of the same number of arms
$f$ as a function of $a$ and $f$. The exponent increases with
increasing of $f$ and $a$ for $f\geq 3$. Fig. 6 presents
$\Theta_{ff'}$ for a fixed $a$ (we have chosen $a=2.7$ for an
illustration). For fixed $f,f'$, this exponent decreases with the
decrease of the correlation parameter $a$. Thus, we can conclude,
that polymer stars interact more weakly in media with
strong correlated disorder.

\section {Conclusions}

The present study provides numerical estimates for the
spectrum of critical exponents that govern the scaling behavior of
the $f$-arm star polymers in a good solvent in the presence of
a correlated disordered medium, characterized by a correlation 
function $g(r)\sim r^{-a}$
at large distances $r$. This extends previous results
\cite{Blavatska-2001b,Blavatska-2001a,Blavatska-2002a}
that have shown that the scaling behavior of polymer chains in
 this type of disorder
belongs to a new universality
class.

Working within the field-theoretical RG approach, we applied both
a double expansion in $\varepsilon=4-d$ and $\delta=4-a$
as well as a technique that evaluates the perturbation series
for fixed $d,a$. The first one-loop analysis allowed us to identify
a quantitatively new behavior in comparison with the pure case.
The second approach, refined by a resummation of the resulting 
divergent series, resulted in numerical quantitative estimates for
the scaling exponents.
We found the numerical values of the exponents $\gamma_f$ in 
three-dimensional case for
different fixed values of the correlation parameter  $2.3\leq
a\leq 2.9$, and for fixed numbers of arms $f=1,\ldots,9$. Depending
on the value of $f$, we find two different regimes of the
entropy-induced effects on the polymer in a correlated porous medium.
While an increase of the correlation of the disorder 
causes the 
free energy of {\em chain} polymers ($f=1,f=2$) to decrease,
the same change in correlation rather leads to an increase in 
the free energy 
for {\em star} polymers ($f\geq 3$). Therefore, for a mixture 
of chain and star polymers of equal molecular mass (same total
number of effective monomers) in a solution for which
a part of the volume is influenced by a porous medium
the disorder-influenced
part of the solvent is predicted to be enriched by chain polymers. 
Correspondingly, the relative concentration of star polymers to
and chain polymers will be lower in the porous medium. 

From our numerical estimates for contact exponents $\Theta_{ff'}$,
we deduce the influence of the correlated disorder for the 
effective interaction between star polymers. Again we find
different behavior for chain and star polymers. While for
chain polymers the effective contact interaction increases
for decreasing $a$, i.e. for enhanced correlation, the mutual
interaction between star polymers is weakened in correlated
media.

\section*{Acknowledgments} \label{A}

We thank Myroslav Holovko, Yurij Kalyuzhnyi, Carlos V\'asquez,
and Ricardo Paredez for discussions.  
The authors acknowledge support by the following
institutions: Alexander von Humboldt foundation (V.B.), European
Community under the Marie Curie Host Fellowships for the Transfer of
Knowledge, project COCOS, contract No. MTKD-CT-2004-517186 (C.v.F.),
and Austrian Fonds zur F\"orderung der wi\-ssen\-schaft\-li\-chen
Forschung under Project No. P16574 (Yu.H.).

\section{Appendix}
Here, we present the expressions for the loop integrals, as they appear
in the RG functions. We make the couplings dimensionless by 
redefining $u=u\hat\mu^{d-4}$ and
$w=w\hat\mu^{a-4}$. Therefore, the loop integrals do not explicitly 
contain the mass.
\begin{eqnarray}
&&I_1=\int\frac{{\rm d}{\vec q}}{(q^2+1)^2};\nonumber\\
&&I_2=\int\frac{{\rm d}{\vec q}\,q^{a-d}}{(q^2+1)^2};\nonumber\\
&&I_3=\int\frac{{\rm d}{\vec q}\,q^{2(a-d)}}{(q^2+1)^2};\nonumber\\
&&I_4=\int\int\frac{{\rm d}{\vec q_1}{\rm d}{\vec
q_2}\,q_1^{(a-d)}}{(q_2^2+1)^2((q_1-q_2)^2+1)^2};\nonumber\\
&&I_5=\int\int\frac{{\rm d}{\vec q_1}{\rm d}{\vec
q_2}\,q_1^{(a-d)}\,q_2^{a-d}}{(q_2^2+1)^2((q_1-q_2)^2+1)^2};\nonumber\\
&&I_6=\int\int\frac{{\rm d}{\vec q_1}{\rm d}{\vec
q_2}}{(q_1^2+1)(q_2^2+1)^2((q_1-q_2)^2+1)^2};\nonumber\\
&&I_7=\int\int\frac{{\rm d}{\vec q_1}{\rm d}{\vec
q_2}\,q_1^{a-d}}{(q_1^2+1)(q_2^2+1)^2((q_1-q_2)^2+1)};\nonumber\\
&&I_8=\int\int\frac{{\rm d}{\vec q_1}{\rm d}{\vec
q_2}\,q_1^{a-d}q_2^{(a-d)}}{(q_1^2+1)(q_2^2+1)^2((q_1-q_2)^2+1)};\nonumber\\
&&I_9=\int\int\frac{{\rm d}{\vec q_1}{\rm d}{\vec
q_2}\,q_1^{a-d}}{(q_1^2+1)^2(q_2^2+1)((q_1-q_2)^2+1)};\nonumber\\
&&I_{10}=\int\int\frac{{\rm d}{\vec q_1}{\rm d}{\vec
q_2}\,q_1^{2(a-d)}}{(q_1^2+1)^2(q_2^2+1)((q_1-q_2)^2+1)};\nonumber\\
&&D_1=\frac{\partial }{\partial k^2}\left [ \int\frac{
{\rm d}\vec{q}\, q^{a-d}}{[q+k]^2 + 1) }\right ]_{k^2=0}.
\end{eqnarray}
The correspondence of the integrals to the diagrams in 
Fig. \ref{diagram} is:
(b): integrals $I_1,I_2,I_3$, $(c)$:  $I_4,I_5$;
$(d)$: $I_1I_2$, $I_2^2$;  $e$: $I_6,I_7,I_8,I_9,I_{10}$;
$(f)$: $I_6,I_7,I_8,I_9$.
In our calculations, we use the following formulas for folding many
denominators into one (see e.g. \cite{Amit-1984B}):
\begin{widetext}
\begin{eqnarray}
\lefteqn{
\frac{1}{a_1^{\alpha_1}\ldots
a_n^{\alpha_n}}=\frac{\Gamma(\alpha_1{+}\ldots{+}\alpha_n)}
{\Gamma(\alpha_1)\ldots
\Gamma(\alpha_n)}
}&&
\nonumber\\
&&
\int\limits_0^1{\rm d}x_1\!\!\ldots\!\! \int\limits_0^1{\rm
d}x_{n-1}
\frac{ x_1^{\alpha_n-1}\ldots x_{n-1}^{\alpha_{n-1}-1}
(1{-}x_1{-}\ldots{-}x_{n-1})
^{\alpha_n-1}}
{(x_1a_1{+}\ldots{+}x_{n-1}a_{n-1}{+}(1{-}x_1{-}\ldots{-}x_{n-1})a_n)^
{\alpha_1+\ldots+\alpha_n}}
\label{param}
\end{eqnarray}
\end{widetext}
To compute the $d$-dimensional integrals we apply
\begin{eqnarray}
&&\int\limits_0^{\infty}
\frac{ {\rm d}q\, q^{d-1}
} {(q^2+2\vec{k}\vec{q}+m^2)^{\alpha}}=
\nonumber\\&&
\frac{1}{2}\frac{\Gamma(d/2)\Gamma(\alpha-d/2)} {\Gamma(\alpha)}
(m^2-k^2)^{d/2-\alpha}\label{int}
\end{eqnarray}

As an example we present the calculation of the integral $I_7$.
First, we make use of formula (\ref{param}) to rewrite:
\begin{eqnarray}
&&\frac{1} {(q_1^2+1)((q_1-q_2)^2+1)}=
\nonumber\\
&&\frac{\Gamma(2)}{\Gamma(1)\Gamma(1)}
\int_0^1\frac{ {\rm d}x } {(q_1^2+2x\vec{q_1}\vec{q_2}+xq_2^2+1)^2}.
\end{eqnarray}
Now one can perform integration over $q_1$, passing to the
$d$-dimensional polar coordinates and making use of the formula
(\ref{int}):
\begin{eqnarray}
&&\int\frac{{\rm d}\vec{q_1}\, q_1^{a-d}}
{(q_1^2+2x\vec{q_1}\vec{q_2}+xq_2^2+1)^2}= \nonumber\\ &&C
\int\limits_0^{\infty}\frac{{\rm d}q_1\, q_1^{a-1}}
{(q_1^2+2x\vec{q_1}\vec{q_2}+xq_2^2+1)^2}=\nonumber\\
&&\frac{1}{2} \frac{\Gamma(a/2)\Gamma(2-a/2)}
{\Gamma(2)}(1+q_2^2x(1-x))^{a/2-2},\end{eqnarray} where the constant
$C=(2\pi)^{d/2}/\Gamma(d/2)$ results from integration over the
angular variables. It does not appear explicitly in the following
expressions. Finally, we are left with:
\begin{eqnarray}
I_7&=&\frac{1}{2} {\Gamma(a/2)}{\Gamma(2-a/2)}
\nonumber\\
&&
\int_0^{\infty}\frac{{\rm d}
q_2\,q_2^{d-1}}{(q_2^2+1)^2}\int\limits_0^1{\rm
d}x(1+q_2^2x(1-x))^{a/2-2},
\end{eqnarray}
this integral was calculated numerically, fixing the values of the
parameters $d, a$ using the MAPLE package.
Note that some of the integrals can also be evaluated analytically.

\begin{widetext}
\begin{center}
\begin{table}
\begin{tabular}
{|c| c  c  c
c c  c c c c c c|}
\hline $a$& $I_1$  &  $I_2 $ & $I_3 $  & $ I_4 $ & $I_5$ & $ I_6
$&$I_7$& $I_8$ &$I_9$& $I_{10}$ & $D_1$
\\ \hline
2.9 & 0.7855 &  0.7155 &  0.6605 &  0.5621 &  0.5119 &  0.4114 &
0.3643 & 0.3477 &  0.3916 &  0.3756 &  0.0052  \\ 2.8 & 0.7855 &
0.6605 & 0.5825 &  0.5187 &  0.4363 &  0.4114 &  0.3274 &  0.3016
& 0.3756 & 0.3525 &  0.0012  \\ 2.7 &  0.7855 &  0.6170 & 0.5345 &
0.4848 &
0.3807 & 0.4114 &  0.2981 &  0.2677 &  0.3626 &  0.3395 & 0.0015  \\
2.6 & 0.7855 &  0.5825 &  0.5085 &  0.4575 & 0.3393 & 0.4114 &
0.2746 & 0.2425 &  0.3525 &  0.3357 & 0.0021  \\ 2.5 & 0.7855 &
0.5550 & 0.5000 & 0.4361 &  0.3080 &  0.4114 &  0.2555 & 0.2238 &
0.3448 & 0.3408 & 0.0027 \\ 2.4 &  0.7855 & 0.5345 & 0.5085 &
0.4198 &
0.2857 & 0.4114 &  0.2406 &  0.2106 &  0.3395 & 0.3557 & 0.0034  \\
2.3 & 0.7855 &  0.5185 &  0.5345 & 0.4075 & 0.2688 & 0.4114 &
0.2283 & 0.2014 &  0.3365 &  0.3823 &  0.0041  \\ \hline
\end{tabular}
\caption{The numerical values for loop integrals $I_i$ for $d=3$ and
different values of the correlation parameter $a$.\label{numer}}
\end{table}
\end{center}
\end{widetext}

Below, we list the results for all the integrals.
\begin{eqnarray}
I_1&=&\frac{1}{2}\Gamma(d/2)\Gamma(2-d/2),\nonumber\\
I_2&=&\frac{1}{2}\Gamma(a/2)\Gamma(2-a/2),\nonumber\\
I_3&=&\frac{1}{2}\Gamma((2a-d)/2)\Gamma(2-(2a-d)/2),\nonumber\\
I_4&=&\frac{1}{4}\Gamma(a/2)\Gamma(d/2)\Gamma(2-a/2)\Gamma(2-d/2),\nonumber\\
I_5&=&\frac{1}{4}\Gamma(a/2)\Gamma(a/2)\Gamma(2-a/2)\Gamma(2-a/2),\nonumber\\
I_6&=&\frac{1}{2} {\Gamma(d/2)}{\Gamma(2-d/2)} \times\nonumber\\
&& \int_0^{\infty}\frac{{\rm d}
q_2\,q_2^{d-1}}{(q_2^2+1)^2}\int\limits_0^1{\rm
d}x(1+q_2^2x(1-x))^{d/2-2},\nonumber\\ I_8&=&\frac{1}{2}
{\Gamma(a/2)}{\Gamma(2-a/2)} \times\nonumber\\ &&
\int_0^{\infty}\frac{{\rm d}
q_2\,q_2^{a-1}}{(q_2^2+1)^2}\int\limits_0^1{\rm
d}x(1+q_2^2x(1-x))^{a/2-2},\nonumber\\ I_9&=&\frac{1}{2}
{\Gamma(d/2)}{\Gamma(2-d/2)} \times\nonumber\\ &&
\int_0^{\infty}\frac{{\rm d}
q_2\,q_2^{a-1}}{(q_2^2+1)^2}\int\limits_0^1{\rm
d}x(1+q_2^2x(1-x))^{d/2-2},\nonumber\\
I_{10}&=&\frac{1}{2}
{\Gamma(d/2)}{\Gamma(2-d/2)} \nonumber\\ &&
\int_0^{\infty}\frac{{\rm d}
q_2\,q_2^{2a-d-1}}{(q_2^2+1)^2}\int\limits_0^1{\rm
d}x(1+q_2^2x(1-x))^{d/2-2},\nonumber\\ D_1&=&
\frac{(a-2)(a-3)(a-4)}{192\sin(\pi(1/2a-1))}.\nonumber
\end{eqnarray}
Note, that the analytical value for $D_1$ is taken from
\cite{Prudnikov-1999Ia,Prudnikov-1999I,Prudnikov-2000I}.
\bibliographystyle{apsrev}
\bibliography{ferholI,ferholB,ferhol,cvfpublist,starlr}
\end{document}